\documentclass[twocolumn,floats,amsmath,amssymb]{revtex4}

\usepackage{epsfig}

\usepackage{graphicx}
\usepackage{dcolumn}
\usepackage{bm}
\usepackage{chemarr}
\usepackage{multirow}
\usepackage{subfigure}
\usepackage{epstopdf}

\usepackage[T1]{fontenc}
\usepackage{color}
\usepackage[english]{babel}
\usepackage{hyperref}
\usepackage{verbatim} 
\usepackage{bbold}
\usepackage{pdfsync}
\usepackage{paralist}
\usepackage{MnSymbol}
\usepackage[version=3]{mhchem}
\usepackage{chemarr}
\usepackage{amsthm}

\DeclareGraphicsExtensions{.png,.gif,.jpg,.pdf,.bmp}

\begin{document}

\title{Noise-induced multistability in chemical systems: Discrete vs Continuum modeling} 

\author{Andrew Duncan $^{1}$, Shuohao Liao $^{1}$, Tomas Vejchodsky $^{1}$, Radek Erban $^{1}$, Ramon Grima $^{2}$}

\affiliation{$^{1}$ Mathematical Institute, University of Oxford,
Radcliffe Observatory Quarter, Woodstock Road,
Oxford, OX2 6GG, U.K., $^{2}$ School of Biological Sciences, Kings Buildings, Mayfield Road, University of Edinburgh, EH9 3JF, U.K.}

\begin{abstract}
The noisy dynamics of chemical systems is commonly studied using either the chemical master equation (CME) or the chemical Fokker-Planck equation (CFPE). The latter is a continuum approximation of the discrete CME approach. We here show that the CFPE may fail to capture the CME's prediction of noise-induced multistability. In particular we find a simple chemical system for which the CME's marginal probability distribution changes from unimodal to multimodal as the system-size decreases below a critical value, while the CFPE's marginal probability distribution is unimodal for all physically meaningful system sizes.  
\end{abstract}

\maketitle

The dynamics of chemical and biochemical circuits is noisy whenever the number of molecules of at least one chemical species is small \cite{vanKampen}. The analysis of such circuits typically proceeds either via the chemical master equation (CME) or the chemical Fokker-Planck equation (CFPE) \cite{vanKampen}. The latter is obtained from a truncation of the Kramers-Moyal expansion of the CME up to second-order derivatives and is hence regarded as a continuous approximation or an asymptotic representation of the CME \cite{Horsthemke1977}. For systems composed of only unimolecular reactions, the CFPE's prediction for the mean and variance of the molecule numbers is well known to be the same as those of the CME \cite{McQuarrie1967}; for systems in which at least one reaction is bimolecular, the CFPE's predictions for the first two moments are not exact but it has recently been shown that the accuracy is high over a wide range of molecule numbers \cite{Grima2011,Schnoerr2014}. 

In the limit of large system sizes, the CFPE can capture noise-induced phenomena which are predicted by the CME (see for example recent studies of noise-induced oscillations \cite{Erban2009,Thomas2013}). A broader question is whether the CFPE can capture noise-induced phenomena which appear at intermediate system-sizes. Of particular interest to us here is the phenomenon of noise-induced multimodality (also known as noise-induced multistability or stochastic multistability) where the marginal probability distribution of the CME switches from a unimodal to a multimodal distribution as the system-size (the volume for chemical systems) is decreased below a critical value. The existence of such a phenomenon has been long known \cite{HortshemkeLevefer} yet its practical relevance to biology and ecology has only started to be appreciated in the past decade \cite{Kepler2001,Qian2009,Thomas2014,Biancalani2014}. Recent evidence suggests that the CFPE does capture the onset of noise-induced bistability (noise-induced bimodality in our terminology) \cite{Biancalani2014}. Here we demonstrate, by means of a biochemically relevant example, that this is not generally the case; in particular we show that while the CME predicts a change from a unimodal to a multimodal distribution as the system-size decreases below a certain value, the CFPE shows a unimodal distribution independent of the system-size. 

{\emph{The chemical system}}. We start by considering the following simple model of transcription regulation without feedback \cite{Kepler2001}:

\begin{equation}
\label{eq:modelG}
	G \underset{k_{off}}{\overset{k_{on}}{\rightleftharpoons}} G^*, \quad
	G \xrightarrow{k_1} G + P, \quad
	G^* \xrightarrow{k_2} G^* + P, \quad
	P  \xrightarrow{c} \emptyset.
\end{equation}

A gene can be in one of two states $G$ and $G^*$; the switching between these two states is random and each state is associated with a different rate of protein ($P$) formation. The protein once formed can also decay. This model is a simplification of more realistic gene models where the mRNA is explicitly modelled \cite{ShahrezaeiSwain2008}. We consider the case where there are $N$ genes such that the total number of $G$ and $G^*$ equals $N$ at all times; although genes typically exist in one or two copies per cell, plasmids are nowadays commonly used to genetically engineer cells with a large number of copies of a given gene \cite{Alberts1994} and hence our model is of biochemical relevance. 

{\emph{Stochastic mesoscopic descriptions of the chemical system}}. Denoting by $t$ the time and by $\tau=ct$ the dimensionless time, the CME for for the reaction system (\ref{eq:modelG}) is given by:
\begin{equation}
\label{eq:cme_nondim}
\begin{aligned}
	\partial_{\tau} \Pi(n, p, \tau) = \, &q_{off} (E_n^{-1} - 1)(N - n)\Pi(n, p, \tau) \\
							 + \,&q_{on} (E_n^{+1} - 1) n \Pi(n, p, \tau) \\
							 + \,&q_1 n (E_p^{-1} - 1) \Pi(n, p, \tau) \\
							 + \,&q_2(N-n) (E_p^{-1} - 1)\Pi(n, p, \tau) \\ 
							 + \,&(E_p^{+1} - 1)p \Pi(n, p, \tau),
\end{aligned}
\end{equation}
where $\Pi(n, p, \tau)$ is the probability that at time $\tau$ there are $n$ genes in state $G$ and $p$ protein molecules, and $E^m_n$ and $E^m_p$ are step operators such that when they act on a function $f \equiv f(n,p)$, their action is $E^m_n f(n,p) = f(n+m,p)$ and $E^m_p f(n,p) = f(n,p+m)$. The non-dimensional reaction rates are given by:
\begin{equation}
        \nonumber
	q_{off} = \frac{k_{off}}{c}, \quad q_{on} = \frac{k_{on}}{c}, \quad q_{1} = \frac{k_{1}}{c}, \quad q_{2} = \frac{k_{2}}{c}.
\end{equation}

The CFPE for the reaction system (\ref{eq:modelG}) (with dimensionless time units, $\tau = ct$) is given by:
\begin{equation}
\label{eq:cfpe}
\begin{aligned}
	\partial_{\tau} \Pi(n,p,\tau) = &-\partial_{n}\left(\left(q_{off}(N-n) - q_{on} n\right) \Pi (n,p,\tau)\right)\\
	    &-\partial_{p}\left(\left(q_{1}n + q_{2}(N -n) - p\right)\Pi(n,p,\tau)\right) \\
	    &+\frac{1}{2}\partial_{n}^2 \left(\left(q_{off}(N-n) + q_{on}n\right) \Pi(n, p,\tau)\right)\\
	    &+\frac{1}{2}\partial_{p}^2\left(\left(q_{1}n + q_{2}(N - n) + p\right)\Pi(n,p,\tau)\right).
\end{aligned}
\end{equation}

{\emph{The quasi-stationary approximation (QSA) of the CME}}. Next we solve the CME under the condition that the timescales governing the decay of small fluctuations about the steady-state mean number of molecules of $G$ and $P$ are well separated. Since the system (\ref{eq:modelG}) consists of purely first-order reactions, the equations for the means $\langle n \rangle$ and $\langle p \rangle$ as obtained from the CME Eq. (\ref{eq:cme_nondim}) are precisely given by the conventional rate equations: $\partial_{\tau} \langle n \rangle = -q_{on} \langle n \rangle + q_{off} (N -  \langle n \rangle)$ and $\partial_{\tau} \langle p \rangle = q_{1} \langle n \rangle + q_{2} (N - \langle n \rangle) - \langle p \rangle$. The two characteristic non-dimensional timescales are given by the absolute value of the inverse of the eigenvalues of the Jacobian of the above system of rate equations and are: $\tau_G = (q_{on} + q_{off})^{-1}$ and $\tau_P = 1$, where the former governs the decay of small fluctuations in $G$ and the latter the same but for $P$. Clearly gene switching between the two states $G$ and $G^*$ is much slower than the rest of the processes in the system whenever $\tau_G \gg 1$, i.e., the protein reaches steady-state in a time much shorter than the time it takes for a gene to switch from one state to another. 

Now we approximately solve the CME Eq. (\ref{eq:cme_nondim}) in the quasi-stationary limit given by $\varepsilon = \tau_G^{-1} \ll 1$. Rescaling time by $\tau \rightarrow \varepsilon \tau$ we obtain the following master equation:
\begin{equation}
\label{eq:cme_singular}
\begin{aligned}
	\partial_{\tau} \Pi(n, p, \tau) = \frac{1}{\varepsilon}\mathcal{L}^0 \Pi(n, p, \tau) + \mathcal{L}^1 \Pi(n, p, \tau),
\end{aligned}
\end{equation}
where the two operators $\mathcal{L}^0$ and $\mathcal{L}^1$ are given by:
\begin{align}
        \label{op1}
	\mathcal{L}^0  &= (q_1 n + q_2(N-n)) (E_p^{-1} - 1) + (E_p^{+1} - 1)p, \\
	\mathcal{L}^1 &= \alpha (E_n^{+1} - 1) n + \beta (E_n^{-1} - 1)(N - n),
\end{align}
and $\alpha = q_{on}/(q_{on} + q_{off})$ and $\beta = 1 - \alpha$. Note that $\mathcal{L}^0$ acts only on the protein numbers $p$ while $\mathcal{L}^1$ acts only on the gene numbers $n$.  

In order to solve Eq. (\ref{eq:cme_singular}) in the limit of small $\varepsilon$, we consider the perturbation ansatz:
\begin{equation}
\label{eq:ansatz} \nonumber
\Pi(n, p) = \Pi_0(n, p) + \varepsilon \Pi_1(n, p) + \ldots + \varepsilon^i \Pi_i(n,p) + \ldots
\end{equation}
Substituting the latter in Eq. (\ref{eq:cme_singular}) and comparing coefficients of powers of $\varepsilon$ we obtain the following equations:
\begin{align} \label{eqO1e}
	& O\biggl(\frac{1}{\varepsilon}\biggr) :  \mathcal{L}^0 \Pi_0 = 0, \\
	& O(1) : \mathcal{L}^0 \Pi_1+ \mathcal{L}^1 \Pi_0 = \partial_{\tau} \Pi_0 = 0.
\end{align} 
Note that $\partial_{\tau} \Pi_0 = 0$ by the assumption of steady-state. By Bayes' theorem we can write  $\Pi_0(n, p) = \Pi_0(p \, | n)\mu(n)$, where $\Pi_0(p | n)$ is the stationary density for $P$ given the number of $G$ molecules is $n$, and $\mu(n)$ is the marginal stationary distribution for $G$. Summing the $O(1)$ equation over $p$, we obtain:
\begin{equation}
\label{summedover} 
\sum_{p \in \mathbb{N}}\left[\mathcal{L}^0 \Pi_1(n, p)\right] + \mathcal{L}^1 \left[\sum_{p\in \mathbb{N}}\Pi_0(p \, | \, n)\mu(n)\right] = 0.
\end{equation} 
The first term on the left hand side is zero by the definition of $\mathcal{L}^0$ in Eq. (\ref{op1}). The second term simplifies by the normalisation condition $\sum_{p \in \mathbb{N}}\Pi_0(p \, | \, n) = 1$. Thus Eq. (\ref{summedover}) reduces to:  
\begin{equation} \nonumber
\mathcal{L}^1 \mu(n) = 0,
\end{equation}
which possesses a unique normalised solution given by:
\begin{equation}
\label{eq:marginal_density}
	\mu(n) = \frac{{N \choose n}}{(1 + \lambda)^N} \lambda^n, \quad n \in \{0,...,N\}
\end{equation}
where $\lambda = \beta / \alpha$. From the $O(\frac{1}{\varepsilon})$ equation we obtain:
\begin{equation}
\label{eq:conditioned_eqn} \nonumber
	\mu(n)\,\mathcal{L}^0 \Pi_0(p \,|\, n) = 0.
\end{equation}
This equation can be easily solved yielding the Poissonian:
\begin{equation}
\label{eq:conditional_density}
\Pi_0(p \,|\, n) = \frac{1}{p!}\left({q_1\,n  + q_2\,(N-n)}\right)^pe^{-(q_1 n + q_2(N-n))}.
\end{equation}
Finally multiplying  Eq. (\ref{eq:marginal_density}) and Eq. (\ref{eq:conditional_density}) we obtain by Bayes' theorem the leading order approximation to $\Pi_0(n, p)$ in the limit of small $\varepsilon$; after marginalising over $n$, we obtain the stationary distribution of the protein numbers in the same limit:
\begin{equation}
\label{eq:stat_dist}
\Pi(p) \approx \sum_{n = 0}^{N} {N \choose n}\frac{\lambda^n}{p!}\frac{\left({q_1\,n  + q_2\,(N-n)}\right)^p}{(1 + \lambda)^N} e^{-(q_1 n + q_2(N-n))}.
\end{equation}
In principle a similar quasi-stationary limit can be applied to the CFPE Eq. (\ref{eq:cfpe}). In practice, however, the equation thus obtained cannot be reduced beyond an integral which has no closed form solution; thus in what follows we obtain the stationary distribution of the protein numbers of the CFPE via numerical integration (see later for details). 

{\emph{Number of modes of the approximate stationary distribution of the CME and CFPE approaches}}. Let $\Omega$ be the volume of the compartment in which the chemical reaction network (\ref{eq:modelG}) is confined. Furthermore let $\phi_P = p / \Omega$ be the protein concentration and $\phi_N = N/\Omega$ be the (constant) total gene concentration. We now study the number of modes of the quasi-stationary distribution of $\phi_P$ as a function of $\Omega$; in this thought experiment, the volume $\Omega$ is varied at constant $\phi_N$ such that an increase in volume, necessarily translates into a proportionate increase in the total number of genes (this also implies that the number of proteins increases with the volume). Now the CME distribution of $\phi_P$, is given by $\pi(\phi_P) = \Omega \Pi(\phi_P \Omega)$. This probability distribution solution consists of a superposition of $N + 1$ Poisson distributions; this is since there are $N + 1$ combinations of $N$ molecules which can be in two states. This implies that $\pi(\phi_P)$ is generally multimodal, with at most $N + 1$ modes. Since $N$ increases with $\Omega$, we would expect the modality of $\pi(\phi_P)$ to increase with the volume, if the Poissonians are well separated. On the other hand,  given that the deterministic rate equations of the chemical system are monostable, we also know, by the system-size expansion \cite{vanKampen}, that in the thermodynamic limit of large volumes, the probability distribution $\pi(\phi_P)$ tends to a Gaussian and thus unimodal. Hence the overall picture is that the number of modes of $\pi(\phi_P)$ should increase with $\Omega$ for $\Omega$ less than some critical volume $\Omega^*$ and decrease with increasing volume for $\Omega > \Omega^*$. The smallest volume possible is that for which there is one gene $N = 1$; hence the number of modes in the limit of small volumes is 2. 

\begin{figure}
\centering
\subfigure[\ $\Omega = 1$]{
\includegraphics[width=62mm]{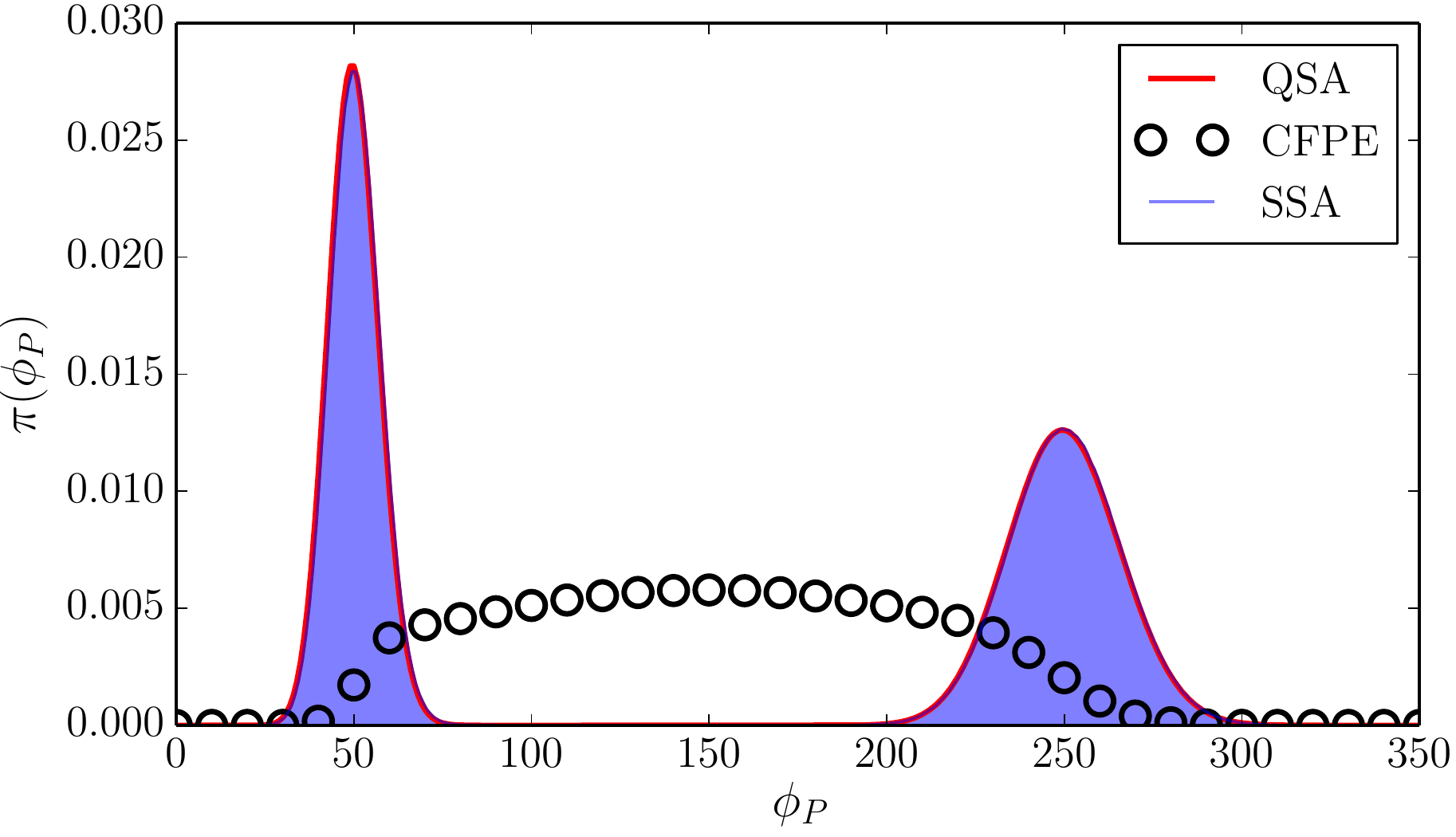}
\label{fig:subfig1}
}
\subfigure[\ $\Omega = 10$]{
\includegraphics[width=62mm]{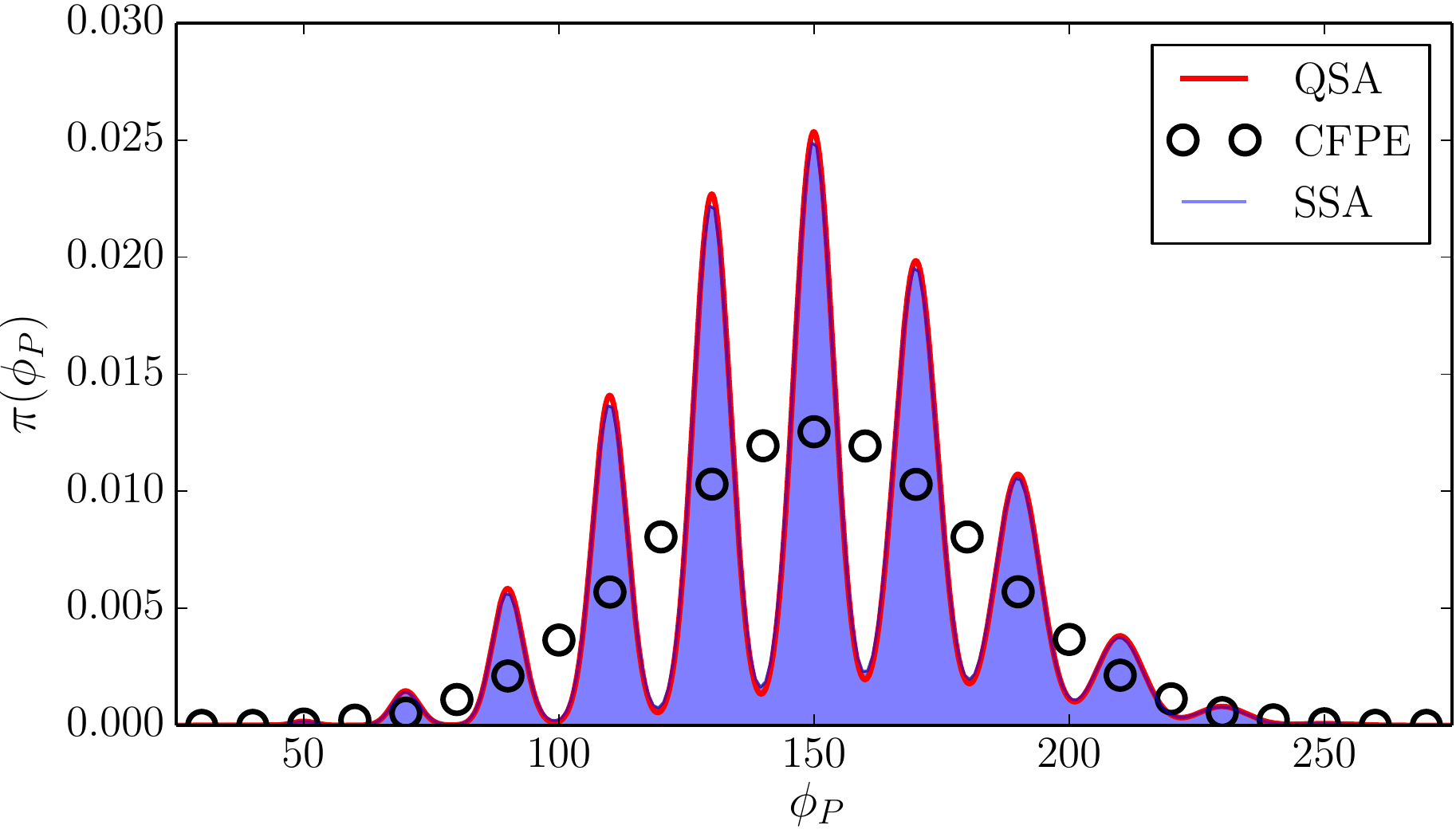}
\label{fig:subfig2}
}
\subfigure[\ $\Omega = 50$]{
\includegraphics[width=62mm]{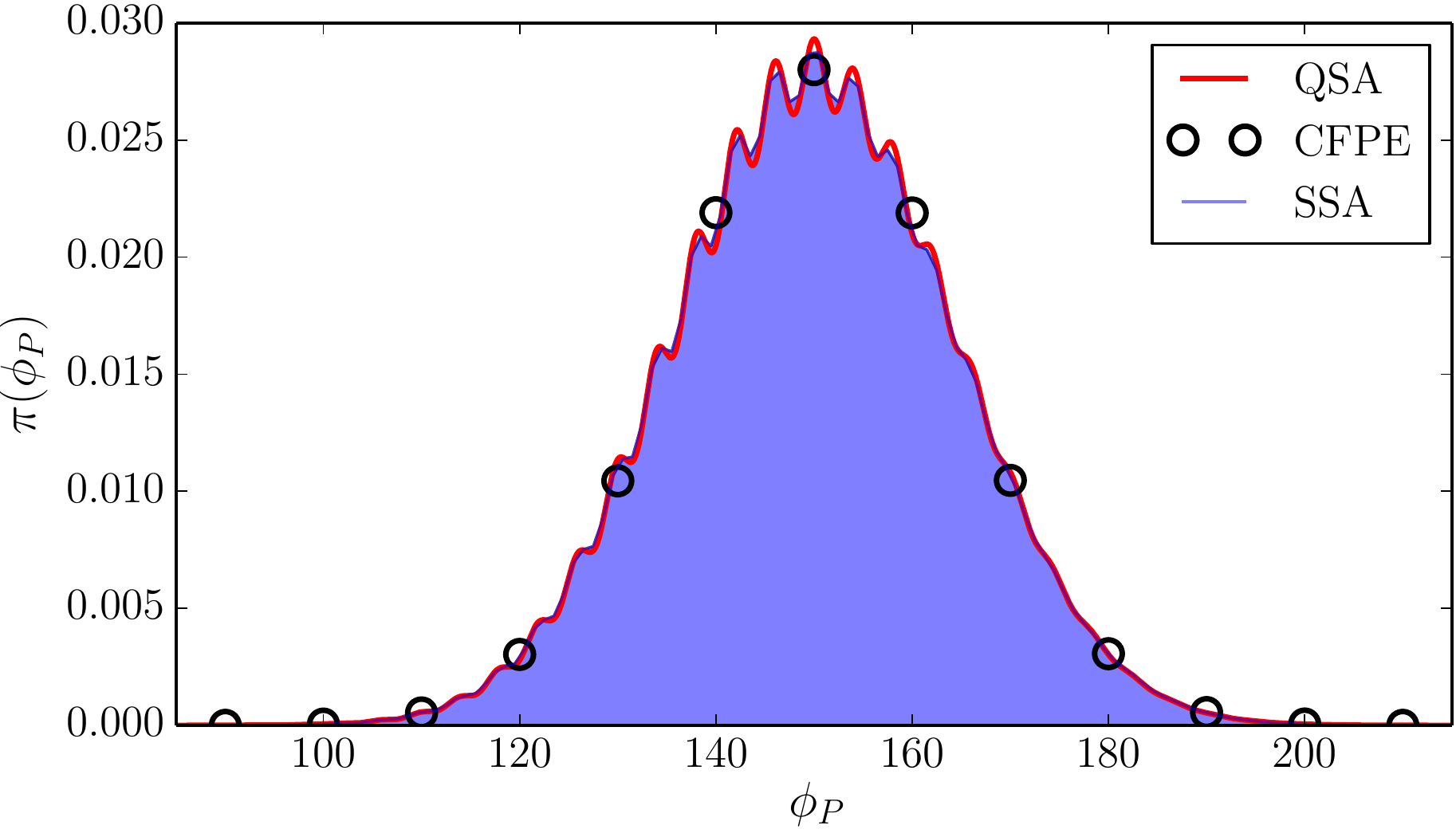}
\label{fig:subfig2}
}
\subfigure[\ $\Omega = 100$]{
\includegraphics[width=62mm]{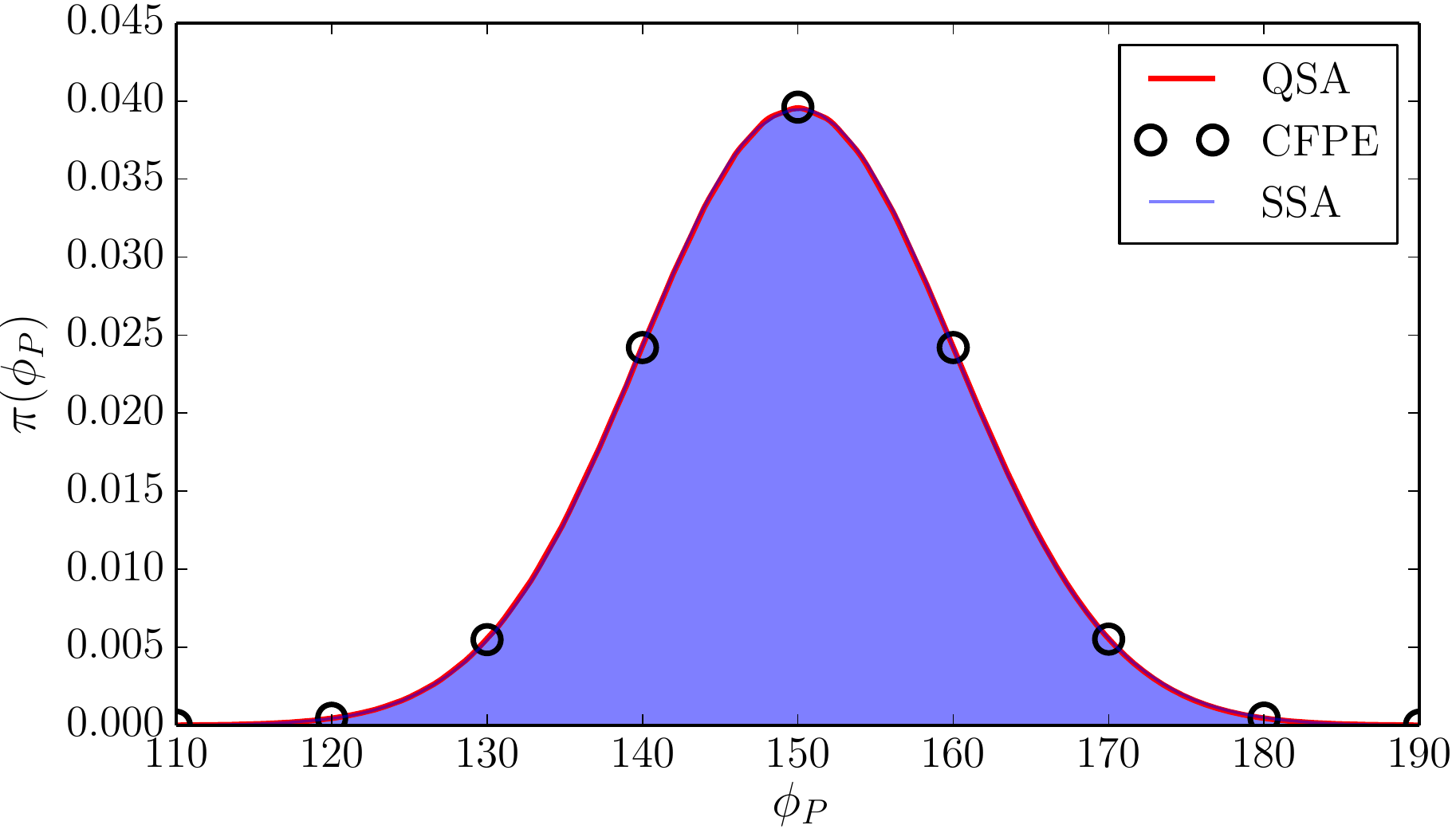}
\label{fig:subfig2}
}
\caption{Comparison of the stationary distribution of protein concentration obtained from SSA simulations with the QSA solution $\pi(\phi_P)$, and the numerical solution of the CFPE for different values of volume, $\Omega$. Parameters (see text) are such that timescale separation is enforced. The QSA and SSA predict multimodality below a certain volume whereas the CFPE predicts a unimodal distribution for all volumes.}\label{fig:plots}
 \end{figure}

In Fig. 1, we plot the QSA solution $\pi(\phi_P)$ for four different volumes with parameters $\phi_N = 1$, $q_{on}=q_{off}=10^{-3}$, $q_1 = 50$, $q_2 = 250$; we compare this with the solution from the CFPE and exact stochastic simulations of the CME using Gillespie's stochastic simulation algorithm (SSA). Note that for the chosen parameters, $\varepsilon = 2 \times 10^{-3} \ll 1$, which implies timescale separation; this is reflected in the excellent agreement between the analytical approximation (QSA) and the SSA. As predicted above, the modality of the probability distribution of the CME goes through a maximum as the volume is progressively increased, with the number of modes for very low and large volumes being 2 and 1, respectively. Thus the CME predicts noise-induced multimodality as the volume is decreased beyond some critical value; we call this ``noise-induced'' since the particle numbers becomes smaller with the volume and the size of intrinsic noise correspondingly increases. The CFPE solution is obtained by discretising the PDE (\ref{eq:cfpe}) using the continuous Galerkin finite element method \cite{Larsson2008} over a triangulation of the domain $[0,N] \times [0, N^*]$ where $N^* > 0$ is some artificial maximum protein number which is chosen sufficiently large such that its value makes no significant difference to the solution; no-flux boundary conditions are imposed along the domain boundaries. Note that the CFPE, unlike the CME, does not predict noise-induced multimodality as the volume is decreased from $100$ to $1$ -- rather it is unimodal for all four volumes in Fig. 1. Note also that for some volumes below $\Omega = 1$, the CFPE does predict two modes of the probability distribution (e.g. a mode at zero and one at a non-zero concentration for $\Omega = 1/2$), however for such volumes the total number of genes is less than 1 (since $\phi_N = 1$); hence we can more precisely state that over the whole range of physically meaningful volumes, the CFPE is unimodal. Note that the no-flux (reflective) boundary conditions on the CFPE are artificial in the sense that unlike the CME, the CFPE does not naturally lead to a restriction of gene numbers between $0$ and $N$ -- the imposition of such artificial reflective conditions can lead to undesirable artefacts in the CFPE predictions (see for example \cite{Schnoerr2014}); we repeated our simulations using the recently developed complex chemical Langevin equation \cite{Schnoerr2014} which avoids the artificial boundary problems and found that for the system studied here, the results obtained are practically the same as shown in Fig. 1. 

\begin{figure} [h]
\centering
\subfigure[\ $q_1 = 2$]{
\includegraphics[width=75mm]{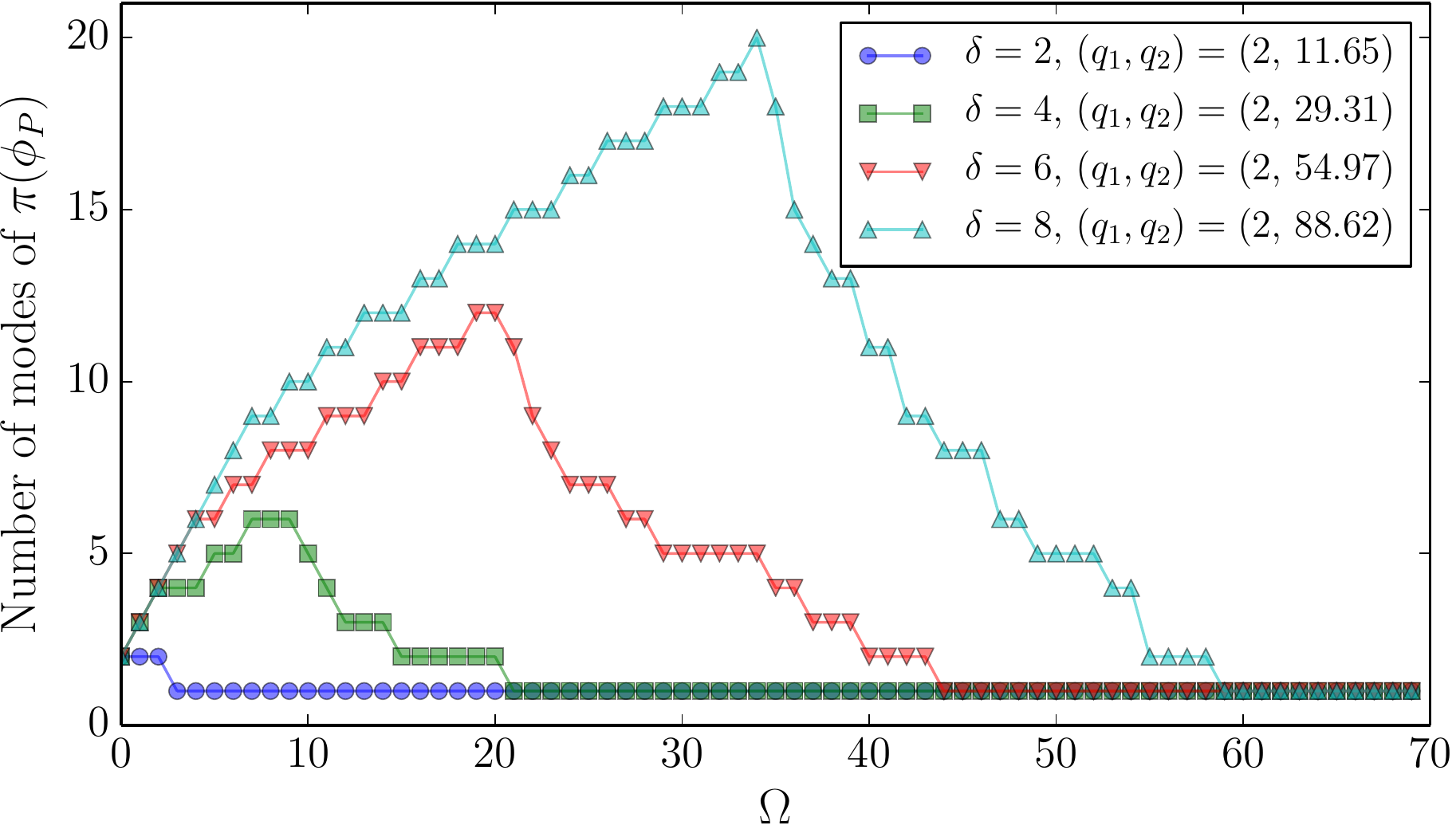}
\label{fig:subfig1}
}
\subfigure[\ $q_1 = 5$]{
\includegraphics[width=75mm]{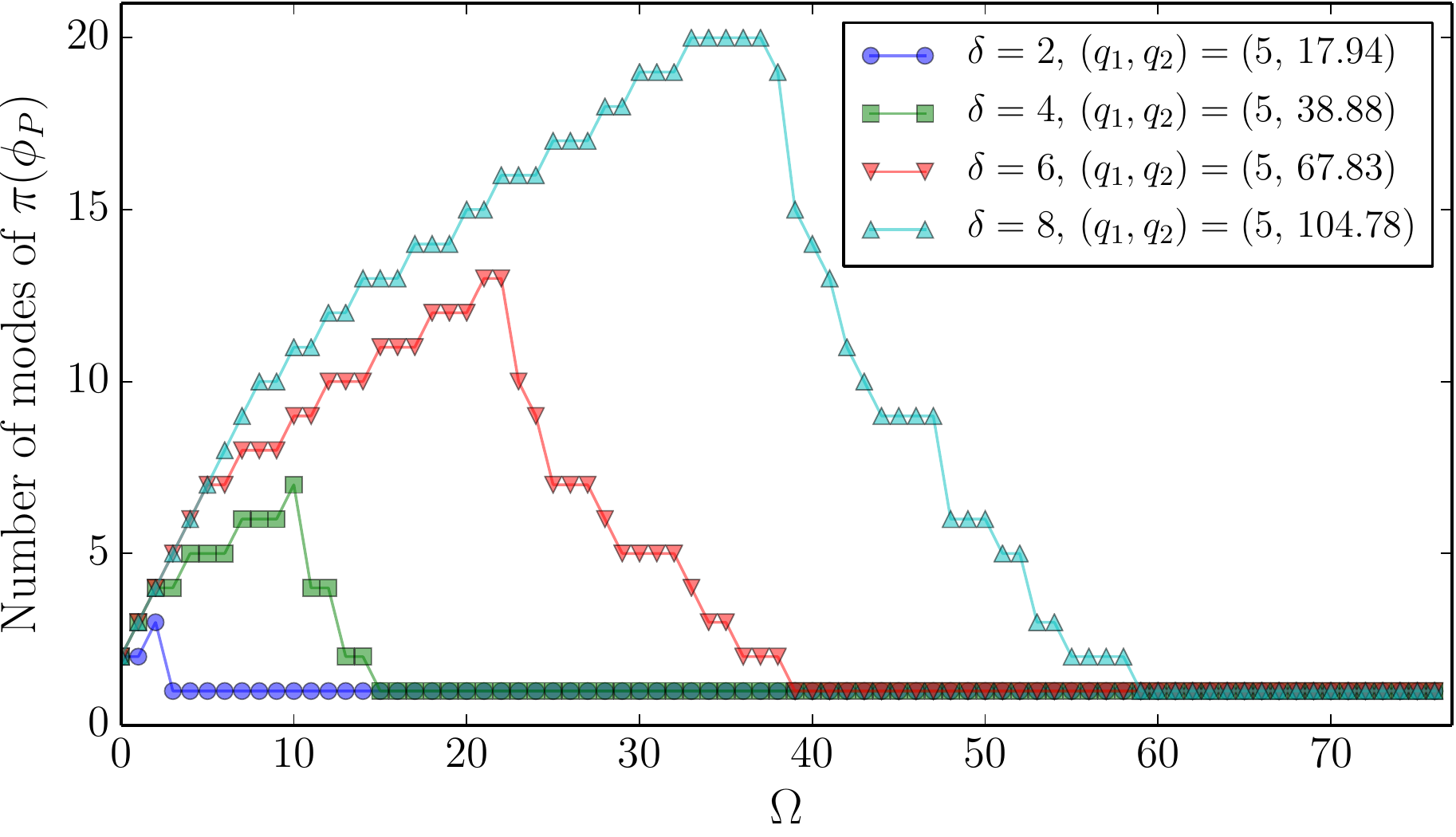}
\label{fig:subfig2}
}
\caption{Plot of the number of modes of $\pi(\phi_P)$ as a function of the volume $\Omega$, for different values of $\delta = \sqrt{q_2} - \sqrt{q_1}$. The larger is the difference between the production rates $q_1, q_2$, the larger is the volume range over which the stationary distribution of the CME is multimodal; in contrast, the solution of the CFPE can be shown to be unimodal for all volumes.}\label{fig:modes}
 \end{figure}

The behaviour elucidated in Fig. 1 is not particular to the parameter set used; in Fig. 2 we show the number of modes of the QSA distribution of protein $\pi(\phi_P)$ as a function of the volume $\Omega$ for 8 different parameter sets (with fixed $\phi_N = 1$, $\lambda=1$). In all cases, the number of modes is $1$ for large volumes, increases as the volume decreases and reaches a maximum at some critical volume; as the volume is decreased further, the number of modes steadily decreases until it reaches a value of 2 at the lowest volume of $\Omega = 1$ (below this volume the total number of genes is less than one and hence unrealistic). The CFPE probability distribution solution is unimodal for all volumes.  A comparison of Fig. 2 (a) and (b) shows that the dimensionless parameter $\delta = \sqrt{q_2} - \sqrt{q_1}$ (and not the individual values of $q_1$ and $q_2$) appears to be the principle factor determining the maximum number of modes of the probability distribution, as well as the critical volume at which this occurs. In particular the larger is $\delta$, the larger is the volume (and the associated number of genes) above which the probability distribution of the CME becomes unimodal and agrees with the CFPE. A heuristic explanation of this is as follows. For the case of one gene ($N = 1$), the QSA solution Eq. (\ref{eq:stat_dist}) predicts two modes, a Poissonian with mean and variance equal to $p = q_1$ and a second Poissonian with mean and variance equal to $p = q_2$. Now say that $q_1 < q_2$; then the condition of well separated Poissonians can be formulated as $q_1 + \sqrt{q_1} \ll q_2 - \sqrt{q_2}$ which is equivalent to $\delta = \sqrt{q_2} - \sqrt{q_1} \gg 1$. The larger is $\delta$, the more pronounced is the bimodal character of the distribution at $N = 1$ and hence one would expect the larger is the total number of genes (and the volume) needed for the multimodality to be washed away -- this is consistent with the role played by $\delta$ in Fig. 2 and which was discussed in the previous paragraph.  

{\emph{Conclusion}}.  We have in this paper shown that while the CFPE leads to mean and variance of fluctuations for the molecule numbers which are the same as the CME for all volumes (since system (\ref{eq:modelG}) is composed of only unimolecular reactions), the CFPE misses the CME prediction of the onset of noise-induced multimodality as the volume is decreased. This is in contrast to that reported in \cite{Biancalani2014} and implies that generally the CFPE does not capture noise-induced multimodality. We note that the probability distribution of the CME Eq. (12) is a superposition of Poissonians and each one is directly associated with an element of the set $\{(0,N),(1,N-1),...,(N,0)\}$ which describes the possible combinations of genes in states $G$ and $G^*$ -- it is this direct association between the distribution and the inherent discreteness of the system that leads to the CFPE's inability to capture the multimodality of the CME. Hence in some sense the phenomenon described in this paper maybe more aptly described as discreteness-induced multimodality. 

Our results also show that the critical volume (and associated gene numbers) above which the probability distribution of the CME is unimodal, as that of the CFPE, increases monotonically with the difference between the two protein production rates; hence although the CFPE is correct in the limit of infinitely large molecule numbers, its breakdown can occur at considerably large molecule numbers if the two protein production rates are well separated. Similar results to the ones reported here are obtained for a chemical system with negative feedback in which the reaction $G \underset{}{\overset{}{\rightleftharpoons}} G^*$ in scheme (\ref{eq:modelG}) is replaced by $G + P\underset{}{\overset{}{\rightleftharpoons}} G^*$. We generally expect the CFPE to fail to reproduce noise-induced multimodality whenever one considers gene regulatory networks under timescale separation conditions (and in a parameter regime where the rate equations are monostable) since in all such cases the probability distribution in the limit of small gene numbers is multimodal and each mode is associated with one of the possible discrete number of gene states \cite{Thomas2014}. 

{\emph{Acknowledgments}}. The research leading to these results has received funding from the European Research Council under the {\it European Community's} Seventh Framework Programme {\it (FP7/2007-2013)} / ERC {\it grant
agreement} No. 239870 (AD, SL, RE). TV acknowledges the support of RVO: 67985840 and from the People Programme (Marie Curie Actions) of the European Union's Seventh Framework Programme (FP7/2007-2013) under REA grant agreement no. 328008. RE would also like to thank the Royal Society for a University Research Fellowship; Brasenose College, University of Oxford, for a Nicholas Kurti Junior Fellowship; and the Leverhulme Trust for a Philip Leverhulme Prize. These prize funds also supported a research visit of RG in Oxford.


\begin{thebibliography}{}
\bibitem{vanKampen} N. G. van Kampen, {\it Stochastic Processes in Physics and Chemistry} (Elsevier, 2007)
\bibitem{Horsthemke1977} W. Horsthemke and L. Brenig, Z. Physik B {\bf{27}}, 341 (1977)
\bibitem{McQuarrie1967} D. A. McQuarrie, J. App. Prob. {\bf{4}}, 413 (1967)
\bibitem{Grima2011} R. Grima, P. Thomas and A. V. Straube, J. Chem. Phys. {\bf{135}}, 084103 (2011)
\bibitem{Schnoerr2014} D. Schnoerr, G. Sanguinetti and R. Grima, J. Chem. Phys. {\bf{141}}, 024103 (2014)
\bibitem{Erban2009} R. Erban et al., SIAM J. Appl. Math. {\bf{70}}, 984 (2009)
\bibitem{Thomas2013} P. Thomas et al. J. Theor. Biol. {\bf{335}}, 222 (2013)
\bibitem{HortshemkeLevefer} W. Horsthemke and R. Lefever, {\it Noise-Induced Transitions: Theory and Applications in Physics, Chemistry, and Biology} (Springer, 1984)
\bibitem{Kepler2001} T. Kepler and T. Elston, Biophys. J. {\bf{81}}, 3116 (2001)
\bibitem{Qian2009} H. Qian, P-Z Shi and J. Xing, Phys. Chem. Chem. Phys. {\bf{11}}, 4861 (2009)
\bibitem{Thomas2014}  P. Thomas, N. Popovic and R. Grima, Proc. Natl. Acad. Sci. USA {\bf{111}}, 6994 (2014)
\bibitem{Biancalani2014} T. Biancalani, L. Dyson and A. J. McKane, Phys. Rev. Letts. {\bf{112}}, 038101 (2014)
\bibitem{ShahrezaeiSwain2008} V. Shahrezaei and P.S. Swain, Proc. Natl. Acad. Sci. USA {\bf{105}}, 17256 (2008)
\bibitem{Alberts1994} B. Alberts et al., {\it Molecular Biology of the Cell} (Garland Publishing, 1994)
\bibitem{Larsson2008} S. Larsson and T. Vidar, {\it Partial differential equations with numerical methods} (Springer, 2008)
\end{thebibliography}
\end{document}